# On the analysis of the Illumina 450K array data: probes ambiguously mapped to the human genome

Xu Zhang[1,6], Wenbo Mu[2] and Wei Zhang[3,4,5*]

[1]Department of Ecology and Evolution, University of Chicago, Chicago, IL, USA
[2]Department of Bioengineering, University of Illinois at Chicago, Chicago, IL, USA
[3]Department of Pediatrics, University of Illinois at Chicago, Chicago, IL, USA
[4]Institute of Human Genetics, University of Illinois at Chicago, Chicago, IL, USA
[5]University of Illinois Cancer Center, Chicago, IL, USA

[6]Current affiliation: Section of Hematology/Oncology, University of Illinois at Chicago, Chicago, IL, USA

**Correspondence:**
Dr. Wei Zhang
University of Illinois at Chicago
Department of Pediatrics
840 South Wood Street, 1200 CSB (MC856)
Chicago, IL 60612, USA
weizhan1@uic.edu

The newly developed Illumina HumanMethylation450 BeadChip (450K array) (Illumina, Inc., San Diego, California) allows unprecedented genome-wide profiling of DNA methylation at >450,000 CpG and non-CpG methylation sites (Sandoval et al., 2011). Utilizing the 450K array, Philibert and colleagues examined the relationship of recent alcohol intake to genome-wide methylation patterns in lymphoblast DNA samples derived from 165 female subjects participating in the Iowa Adoption Studies (Philibert et al., 2012). The authors' interesting paper demonstrated that the 450K array could be a useful tool for ongoing and newly designed epigenome projects. However, given the unique design of the platform (for detailed annotations for the 450K array including probe sequences: http://www.illumina.com/), some cautions might need to be exerted when analyzing the 450K array data, in addition to some general challenges for analyzing the whole-genome DNA methylation data (Laird, 2010). Particularly, we found that a substantial proportion of the >450,000 DNA methylation probes on the 450K array are not aligned to unique, unambiguous loci in the human genome (Moen et al., 2012). In total, we found ~140,000 methylation probes ambiguously mapped to multiple locations in the human genome (hg19) with up to two mismatches in the probe sequences using Bowtie (v2.0.0 beta2) (Langmead et al., 2009;Langmead and Salzberg, 2012). Briefly, Bowtie is an ultrafast, memory-efficient short read aligner by indexing the genome with an extended Burrows-Wheeler technique, which implements a novel quality-aware backtracking algorithm that permits mismatches (Langmead et al., 2009;Langmead and Salzberg, 2012). Different alignment algorithms, e.g., BLAT (Kent, 2002) and MAQ (Li et al., 2008), would provide similar estimates (unpublished data). In comparison, ~1,000 methylation probes were found to be ambiguously mapped to the human genome hg18 in the earlier 27K Illumina HumanMethylation array (27K array) (Bell et al., 2011). Because the much more comprehensive 450K array covers not only promoters, but also gene bodies, untranslated regions (UTRs) and "open sea" methylation sites, the problem of ambiguous alignment may particularly need to be taken into account when analyzing the data from this new platform. Notably, 20 CpG methylation probes (e.g., cg24023553 in Table II; cg00004209 in Table III; cg24675557 in Table V) (Philibert et al., 2012) out of the 90 top-ranking probes reported by Philibert et al. were mapped to ambiguous loci in the current human reference (hg19) using Bowtie (Langmead et al., 2009;Langmead and Salzberg, 2012). Since the problem of ambiguous alignment to the human genome may cause unreliable measurement of DNA methylation level at a particular methylation site, considering this unique problem for this platform may not only facilitate the data analysis (e.g., by improving the multiple-testing problem by removing those affected probes), but also help interpret the results by focusing on more reliable biological signals. In addition, other factors (e.g., polymorphisms in the target sequences, potential batch effects) that may affect other platforms (e.g., the 27K array) (Bell et al., 2011;Fraser et al., 2012) as well may also need to be considered in the analysis of these data.

**Acknowledgements**

This work was supported, in part, by a grant, R21HG006367 (to WZ) from the NHGRI/NIH. The authors declare no conflicts of interest.